# Comment on "Quantum key distribution with 1.25 Gbps clock synchronization" by J. C. Bienfang et al., quant-ph/0405097

**Richard J. Hughes and Jane E. Nordholt**

*Physics Division, Los Alamos National Laboratory, Los Alamos, NM 87545*
hughes@lanl.gov; jnordholt@lanl.gov

**Abstract:** We analyze the significance for quantum key distribution (QKD) of free-space quantum communications results reported in a recent paper (J. C. Bienfang et al., quant-ph/0405097, hereafter referred to as "Bienfang et al."), who contrast the quantum communications rate of their partial QKD implementation (which does not produce cryptographically useful shared, secret keys) over a short transmission distance, with the secret bit rates of previous full QKD implementations over much longer distances. We show that when a cryptographically relevant comparison with previous results is made, the system described by Bienfang et al. would offer no advantages for QKD, contrary to assertions in their paper and in spite of its high clock rate. Further, we show that the claim made by Bienfang et al. that "high transmission rates serve … to extend the distance over which a QKD system can operate" is incorrect. Our analysis illustrates an important aspect of QKD that is too often overlooked in experiments: the sifted bit rate can be a highly misleading indicator of the performance of a QKD system.

In a recent paper ("Quantum key distribution with 1.25 Gbps clock synchronization," quant-ph/0405097), Bienfang et al. report the results of a quantum communications experiment over a 730-m line-of-sight path at night, at a clock rate of $R_{clock}$ = 312 MHz (one-quarter of the value implied by the title of their paper). This clock rate is a factor of 30 – 300 higher, but the path length shorter by a factor of at least 14, than in previously reported free-space quantum key distribution (QKD) experiments, such as our own [1], with which they contrast their results. Bienfang et al. state that: they "have demonstrated transmission of quantum cryptographic key at rates of 1.0 Mbps; roughly two orders of magnitude faster than reported previously". However, the relevance of these results for QKD is obscure and the significance of the comparison with previous work is misleading, because as noted elsewhere in their paper, Bienfang et al. have only implemented the first part of a full QKD protocol ("sifting"), whereas it is the (necessarily smaller) rate of final secret bit production from a complete QKD protocol implementation, such as in [1], that is the cryptographically significant figure-of-merit. Bienfang et al. also fail to take into account the inevitable fall-off in quantum communications rate with transmission distance, owing to beam diffraction and atmospheric attenuation, in their comparison of their results over a short range with previous, longer range results, such as in [1]. In this comment we show that when a cryptographically relevant comparison with such longer range QKD results [1] is made, the quantum communications system of Bienfang et al. would offer no advantages for QKD in terms of either the secret bit production rate or the maximum cryptographically useful transmission range. We also show that the surprising claim made by Bienfang et al., that "high transmission rates serve … to extend the distance over which a QKD system can operate" is incorrect.

The objective of QKD is for the sender ("Alice") and receiver ("Bob") to produce shared, identical, secret random bit strings, for use as cryptographic keys. A full QKD protocol involves a first, quantum communications and post-selection phase, known as "sifting". This is the extent to which QKD has been implemented by Bienfang et al., but sifted bits are completely unsuitable for Alice and Bob to use as cryptographic keys because they are neither perfectly correlated nor completely secret (from eavesdropper, "Eve"). For example, Bienfang et al. report a 1.1% sifted bit error rate (BER): if Alice were to use 256 of these sifted bits as a cryptographic key to encrypt a message to Bob using NIST's Advanced Encryption Standard (AES) algorithm, the probability that Bob could successfully decrypt Alice's message is less than 6%, because even a single error in Bob's key cannot be tolerated. Worse yet, because of the Poisson statistics of the attenuated laser pulses used by Bienfang et al. (at a mean photon number of µ = 0.15), on average 38 of these sifted bits would have been transmitted by signals containing more than one photon, and hence have no guarantee of being secret. This is why in a full QKD implementation such as in [1], sifting must be followed by a second, classical post-processing phase known as "reconciliation and privacy amplification", not considered in Bienfang et al., in which Alice and Bob distill shorter, shared (identical), secret cryptographic keys from their larger sifted bit sequences [2].

The privacy amplification compression factor involved in distilling secret bits from sifted bits can be considerable. It is even possible that Alice and Bob may be unable to distill any secret bits whatsoever, no matter how fast they can generate sifted bits, if they cannot establish a sufficiently high-quality (low noise and low loss) quantum channel. Such a quantum communications system would clearly be of no use for QKD. Thus the cryptographically relevant figure of merit for the performance of a QKD system is the rate of production of final, shared secret bits, and not the quantum communications (sifted bit) rate quoted by Bienfang et al. A cryptographically relevant comparison of the results of Bienfang et al. with previous, full QKD implementations over much longer ranges can be accomplished using the methodology that we developed in [1].

We denote the probability that a transmitted quantum signal results in a sifted bit by $P_{sift}$. Using the methodology developed in [1] and the well-known result of Yurke [3] we may write

$$P_{sift} \approx 1 - \exp(-\mu\eta) \qquad (1)$$

in the system described in Bienfang et al. where µ < 1 is the mean photon number of the transmitted signals, and η is the combined transmission, detection and sifting efficiency for a photon leaving the launch telescope at the transmitter. Using the values quoted in Bienfang et al. we calculate that at their stated transmission distance of $L$ = 0.73km and mean photon number of µ = 0.15 this probability is

$$P_{sift}(0.73) = 2.84 \times 10^{-3} \qquad (2)$$

(Background photons and detector dark noise make negligible contributions to the sifted bits under the conditions reported in Bienfang et al. over this short range.) This results in a calculated sifted bit rate of

$$R_{sift}(0.73) = R_{clock} \times P_{sift}(0.73) = 890 \ kbps \qquad (3)$$

in good agreement with the 900 kbps calculated value quoted by Bienfang et al., but fully 30% higher than their measured value. We will return to this later.

To perform a relevant comparison with our results over a 10-km range [1] we must first re-scale the sifted bit rate (3). Bienfang et al. report "significant diffraction" of the beam emerging from their transmit aperture, and so we reduce their 0.73-km transmission factor, η, by a geometrical factor of $(0.73/10)^2$ = 0.5% and by a further factor of ~ 0.8 for (molecular) atmospheric attenuation at their quantum channel wavelength of 845nm. (We assume a high desert atmospheric extinction in order to make a fair comparison with the results of [1].) Over such a 10km range, nighttime background photons and detector dark noise counts, reported to be 1 kHz on each detector by Bienfang et al., can no longer be ignored as a source of sifted bits. Of these noise counts, one half will contribute errors to Bob's sifted key. Therefore, we estimate that over a 10-km range similar to that used in our own experiment [1], the system described in Bienfang et al. would be capable of producing sifted bits at a rate of

$$R_{sift}(10) = 4.5 \; kbps \qquad (4)$$

at night, at a mean photon number of $\mu = 0.15$, with a sifted BER of $\varepsilon = 12\%$, taking into account the 1.1% BER reported to be intrinsic to the quantum channel. This sifted bit rate is more than two orders of magnitude smaller than the rate reported over the short transmission distance in Bienfang et al., and the BER is larger by a factor of between five and ten than our reported results at the same mean photon number [1] at night. (The sifted bit rate would be doubled in a BB84 protocol [4], but the sifted BER and analysis below would be unchanged.)

Although the essential privacy amplification stage was not included or analyzed in Bienfang et al., it is possible to estimate their system's potential secret bit rate at a 10 km range as follows. We denote the yield of final secret bits, expressed as a fraction of the number of sifted bits, as $F_{sift \rightarrow secret}$, so that the overall secret bit production efficiency of a QKD session is $P_{secret} = P_{sift} \times F_{sift \rightarrow secret}$. Using as a privacy amplification baseline the same "BBBSS91" model [5] for Eve's eavesdropping strategy adopted in [1], $F_{sift \rightarrow secret}$ may be derived from the intermediate quantity

$$r = 1 - \mu - 4\varepsilon \log_2 1.5 + 1.16\left[\varepsilon \log_2 \varepsilon + (1-\varepsilon)\log_2(1-\varepsilon)\right] \qquad (5)$$

which is closely related to Alice and Bob's collision entropy per sifted key bit [2]. For $r > 0$, Alice and Bob can extract (by universal hashing [6]) approximately $F_{sift \rightarrow secret} = r$ secret bits per sifted bit to form a final, shared secret cryptographic key. (This results in a slight over-estimate in the number of secret bits because the well-known privacy amplification "safety factor" is ignored. Likewise for present purposes we ignore the small cost in secret bits to perform the essential authentication part of the QKD protocol.) But when $r$ is zero or negative, $F_{sift \rightarrow secret} = 0$ and no secret cryptographic key bits can be produced. For instance, this will happen if Bob's sifted key BER is too high: Alice must transmit so much parity information about her sifted key in order for Bob to correct his errors that they cannot establish any shared secret cryptographic key whatsoever, no matter how quickly they may generate sifted bits.

In the above expression (5) the second term represents the reduction in secret bits owing to the portion of signals entering the sifted key that emerged from Alice's transmitter with more than one photon (deemed to be known to Eve), the third term accounts for Eve's expected partial information gain from intercept/resend eavesdropping in the Breidbart basis [5], and the final term accounts for the error correction bits that Alice must communicate to Bob. In this estimate we assume use of the CASCADE interactive error correction protocol [7], with efficiency 16% above the Shannon limit. This is expected to be more efficient than any forward error correcting code that might be used.

For the system described in Bienfang et al. over a 10-km nighttime high desert range we then estimate $r = -4\%$ at a mean photon number of $\mu = 0.15$, showing that even if augmented with the essential classical QKD protocol ingredients, their system would be unable to produce any shared, secret cryptographic key bits under these conditions, in spite of its multi-kbps projected sifted bit rate (4). In contrast, the system described in [1] reliably produced shared, secret bits at rates ~ 200 bps under these conditions, (and even in daylight) even though it operated with a ~ 300 times slower clock rate. Furthermore, the system in [1] was projected to have a maximum cryptographically useful range of 45 km at night. Thus, we find that in spite of the high clock rate, and contrary to the claims made in Bienfang et al, their system offers no advantages for QKD over our own, in terms of either the cryptographically relevant figure of merit (secret bit rate) or increased useful range.

We further emphasize that as the above analysis illustrates, and contrary to the assertion of Bienfang et al., increasing the quantum communications clock rate will not extend the useful range of QKD. Indeed, the maximum cryptographically useful range of a QKD system is reached when no secret bits remain after privacy amplification (i.e. $P_{secret} = 0$), and this range is determined by the mean photon number of the signals, the transmission loss and the sifted bit error rate, but is independent of the clock rate. (See Eqs. (1) and (5).) Therefore, once the maximum range is reached increasing the clock rate will have no effect whatsoever on the production of secret bits: it will remain zero.

Bienfang et al report that their system operates at high sifted bit rates over their short 730-m range at night, which may be relevant for QKD in campus-like environments, as opposed to the focus of the 10-km experiment in [1] on much longer ranges. However, we note that their system has a large, unexplained 30% discrepancy between the calculated and observed sifted bit rates. It is well-known in QKD that in certain circumstances transmission losses can be exploited by an adversary to launch very powerful eavesdropping attacks [5]. Also, Bienfang et al. report that their sifted BER is essentially independent of mean photon number, and they only account for ~ one-quarter of their sifted bit errors, leaving ~ three-quarters of the errors unexplained. No known atmospheric optical phenomenon can account for this discrepancy, suggesting an unrecognized system error source. The absence of explanations for these two system properties, which must be understood for the privacy amplification analysis required to implement QKD, brings into question the suitability of the quantum communications system of Bienfang et al. for QKD even over short ranges.

In conclusion, we have clarified the significance for QKD of the high quantum communications rates over a short transmission range reported by Bienfang et al. We have corrected their comparison between their quantum communications results and those of our own, longer range QKD results [1], placing it on a cryptographically relevant footing by incorporating the impact of the essential privacy amplification ingredient and allowing for the great difference in transmission distances, both ignored by Bienfang et al. We have shown that in spite of its high clock rate, their quantum communications system would offer no advantages over our own full QKD implementation in terms of either the cryptographically relevant secret bit rate or maximum useful range. Further, we have shown that their assertion that "high transmission rates … extend the distance over which a QKD system can operate" is incorrect. We emphasize that, as our analysis illustrates, quantum communications (sifted bit) rates present an overly optimistic and potentially very misleading indicator of a QKD system's performance, for which the cryptographically relevant figure of merit is the secret bit rate.